
\catcode`\@=11
\font\tensmc=cmcsc10      
\def\smc{\tensmc}

\def\hcorrection#1{\advance\hoffset by #1 }
\def\vcorrection#1{\advance\voffset by #1 }
\def\wlog#1{}
\newif\iftitle@
\outer\def\title{\title@true\vglue 24\p@ plus 12\p@ minus 12\p@
   \bgroup\let\\=\cr\tabskip\centering
   \halign to \hsize\bgroup\tenbf\hfill\ignorespaces##\unskip\hfill\cr}
\def\endtitle{\cr\egroup\egroup\vglue 18\p@ plus 12\p@ minus 6\p@}
\outer\def\author{\iftitle@\vglue -18\p@ plus -12\p@ minus -6\p@\fi\vglue
    12\p@ plus 6\p@ minus 3\p@\bgroup\let\\=\cr\tabskip\centering
    \halign to \hsize\bgroup\smc\hfill\ignorespaces##\unskip\hfill\cr}
\def\endauthor{\cr\egroup\egroup\vglue 18\p@ plus 12\p@ minus 6\p@}
\outer\def\heading{\bigbreak\bgroup\let\\=\cr\tabskip\centering
    \halign to \hsize\bgroup\smc\hfill\ignorespaces##\unskip\hfill\cr}
\def\endheading{\cr\egroup\egroup\nobreak\medskip}

\outer\def\endproclaim{\par\ifdim\lastskip<\medskipamount\removelastskip
  \penalty 55 \fi\medskip\rm}
\outer\def\demo#1{\par\ifdim\lastskip<\smallskipamount\removelastskip
    \smallskip\fi\noindent{\smc\ignorespaces#1\unskip:\enspace}\rm
      \ignorespaces}

\newcount\footmarkcount@
\footmarkcount@=1
\def\makefootnote@#1#2{\insert\footins{\interlinepenalty=100
  \splittopskip=\ht\strutbox \splitmaxdepth=\dp\strutbox
  \floatingpenalty=\@MM
  \leftskip=\z@\rightskip=\z@\spaceskip=\z@\xspaceskip=\z@
  \noindent{#1}\footstrut\rm\ignorespaces #2\strut}}
\def\footnote{\let\@sf=\empty\ifhmode\edef\@sf{\spacefactor
   =\the\spacefactor}\/\fi\futurelet\next\footnote@}
\def\footnote@{\ifx"\next\let\next\footnote@@\else
    \let\next\footnote@@@\fi\next}
\def\footnote@@"#1"#2{#1\@sf\relax\makefootnote@{#1}{#2}}
\def\footnote@@@#1{$^{\number\footmarkcount@}$\makefootnote@
   {$^{\number\footmarkcount@}$}{#1}\global\advance\footmarkcount@ by 1 }

\hyphenation{man-u-script man-u-scripts ap-pen-dix ap-pen-di-ces}
\hyphenation{data-base data-bases}
\ifx\amstexloaded@\relax\catcode`\@=13
  \endinput\else\let\amstexloaded@=\relax\fi
\newlinechar=`\^^J
\def\eat@#1{}
\def\Space@.{\futurelet\Space@\relax}
\Space@. %
\newhelp\athelp@
{Only certain combinations beginning with @ make sense to me.^^J
Perhaps you wanted \string\@\space for a printed @?^^J
I've ignored the character or group after @.}
\def\futureletnextat@{\futurelet\next\at@}
{\catcode`\@=\active
\lccode`\Z=`\@ \lowercase
{\gdef@{\expandafter\csname futureletnextatZ\endcsname}
\expandafter\gdef\csname atZ\endcsname
   {\ifcat\noexpand\next a\def\next{\csname atZZ\endcsname}\else
   \ifcat\noexpand\next0\def\next{\csname atZZ\endcsname}\else
    \def\next{\csname atZZZ\endcsname}\fi\fi\next}
\expandafter\gdef\csname atZZ\endcsname#1{\expandafter
   \ifx\csname #1Zat\endcsname\relax\def\next
     {\errhelp\expandafter=\csname athelpZ\endcsname
      \errmessage{Invalid use of \string@}}\else
       \def\next{\csname #1Zat\endcsname}\fi\next}
\expandafter\gdef\csname atZZZ\endcsname#1{\errhelp
    \expandafter=\csname athelpZ\endcsname
      \errmessage{Invalid use of \string@}}}}
\def\atdef@#1{\expandafter\def\csname #1@at\endcsname}
\newhelp\defahelp@{If you typed \string\define\space cs instead of
\string\define\string\cs\space^^J
I've substituted an inaccessible control sequence so that your^^J
definition will be completed without mixing me up too badly.^^J
If you typed \string\define{\string\cs} the inaccessible control sequence^^J
was defined to be \string\cs, and the rest of your^^J
definition appears as input.}
\newhelp\defbhelp@{I've ignored your definition, because it might^^J
conflict with other uses that are important to me.}
\def\define{\futurelet\next\define@}
\def\define@{\ifcat\noexpand\next\relax
  \def\next{\define@@}%
  \else\errhelp=\defahelp@
  \errmessage{\string\define\space must be followed by a control
     sequence}\def\next{\def\garbage@}\fi\next}
\def\undefined@{}
\def\preloaded@{}
\def\define@@#1{\ifx#1\relax\errhelp=\defbhelp@
   \errmessage{\string#1\space is already defined}\def\next{\def\garbage@}%
   \else\expandafter\ifx\csname\expandafter\eat@\string
         #1@\endcsname\undefined@\errhelp=\defbhelp@
   \errmessage{\string#1\space can't be defined}\def\next{\def\garbage@}%
   \else\expandafter\ifx\csname\expandafter\eat@\string#1\endcsname\relax
     \def\next{\def#1}\else\errhelp=\defbhelp@
     \errmessage{\string#1\space is already defined}\def\next{\def\garbage@}%
      \fi\fi\fi\next}
\def\famzero{\fam\z@}

\def\lim{\mathop{\famzero lim}}

\def\log{\mathop{\famzero log}\nolimits}

\def\textfont@#1#2{\def#1{\relax\ifmmode
    \errmessage{Use \string#1\space only in text}\else#2\fi}}
\textfont@\rm\tenrm
\textfont@\it\tenit
\textfont@\sl\tensl
\textfont@\bf\tenbf
\textfont@\smc\tensmc
\let\ic@=\/
\def\/{\unskip\ic@}
\def\textfonti{\the\textfont1 }
\def\t#1#2{{\edef\next{\the\font}\textfonti\accent"7F \next#1#2}}
\let\B=\=
\let\D=\.
\def~{\unskip\nobreak\ \ignorespaces}
{\catcode`\@=\active
\gdef\@{\char'100 }}
\atdef@-{\leavevmode\futurelet\next\athyph@}
\def\athyph@{\ifx\next-\let\next=\athyph@@
  \else\let\next=\athyph@@@\fi\next}
\def\athyph@@@{\hbox{-}}
\def\athyph@@#1{\futurelet\next\athyph@@@@}
\def\athyph@@@@{\if\next-\def\next##1{\hbox{---}}\else
    \def\next{\hbox{--}}\fi\next}
\def\.{.\spacefactor=\@m}
\atdef@.{\null.}
\atdef@,{\null,}
\atdef@;{\null;}
\atdef@:{\null:}
\atdef@?{\null?}
\atdef@!{\null!}
\def\srdr@{\thinspace}
\def\drsr@{\kern.02778em}
\def\sldl@{\kern.02778em}
\def\dlsl@{\thinspace}
\atdef@"{\unskip\futurelet\next\atqq@}
\def\atqq@{\ifx\next\Space@\def\next. {\atqq@@}\else
         \def\next.{\atqq@@}\fi\next.}
\def\atqq@@{\futurelet\next\atqq@@@}
\def\atqq@@@{\ifx\next`\def\next`{\atqql@}\else\def\next'{\atqqr@}\fi\next}
\def\atqql@{\futurelet\next\atqql@@}
\def\atqql@@{\ifx\next`\def\next`{\sldl@``}\else\def\next{\dlsl@`}\fi\next}
\def\atqqr@{\futurelet\next\atqqr@@}
\def\atqqr@@{\ifx\next'\def\next'{\srdr@''}\else\def\next{\drsr@'}\fi\next}

\def\textfontii{\the\textfont2 }
\def\{{\relax\ifmmode\lbrace\else
    {\textfontii f}\spacefactor=\@m\fi}
\def\}{\relax\ifmmode\rbrace\else
    \let\@sf=\empty\ifhmode\edef\@sf{\spacefactor=\the\spacefactor}\fi
      {\textfontii g}\@sf\relax\fi}
\def\nonhmodeerr@#1{\errmessage
     {\string#1\space allowed only within text}}
\def\linebreak{\relax\ifhmode\unskip\break\else
    \nonhmodeerr@\linebreak\fi}
\def\allowlinebreak{\relax
   \ifhmode\allowbreak\else\nonhmodeerr@\allowlinebreak\fi}
\newskip\saveskip@
\def\nolinebreak{\relax\ifhmode\saveskip@=\lastskip\unskip
  \nobreak\ifdim\saveskip@>\z@\hskip\saveskip@\fi
   \else\nonhmodeerr@\nolinebreak\fi}
\def\newline{\relax\ifhmode\null\hfil\break
    \else\nonhmodeerr@\newline\fi}
\def\nonmathaerr@#1{\errmessage
     {\string#1\space is not allowed in display math mode}}
\def\nonmathberr@#1{\errmessage{\string#1\space is allowed only in math mode}}
\def\mathbreak{\relax\ifmmode\ifinner\break\else
   \nonmathaerr@\mathbreak\fi\else\nonmathberr@\mathbreak\fi}
\def\nomathbreak{\relax\ifmmode\ifinner\nobreak\else
    \nonmathaerr@\nomathbreak\fi\else\nonmathberr@\nomathbreak\fi}
\def\allowmathbreak{\relax\ifmmode\ifinner\allowbreak\else
     \nonmathaerr@\allowmathbreak\fi\else\nonmathberr@\allowmathbreak\fi}
\def\pagebreak{\relax\ifmmode
   \ifinner\errmessage{\string\pagebreak\space
     not allowed in non-display math mode}\else\postdisplaypenalty-\@M\fi
   \else\ifvmode\penalty-\@M\else\edef\spacefactor@
       {\spacefactor=\the\spacefactor}\vadjust{\penalty-\@M}\spacefactor@
        \relax\fi\fi}
\def\nopagebreak{\relax\ifmmode
     \ifinner\errmessage{\string\nopagebreak\space
    not allowed in non-display math mode}\else\postdisplaypenalty\@M\fi
    \else\ifvmode\nobreak\else\edef\spacefactor@
        {\spacefactor=\the\spacefactor}\vadjust{\penalty\@M}\spacefactor@
         \relax\fi\fi}
\def\newpage{\relax\ifvmode\vfill\penalty-\@M\else\nonvmodeerr@\newpage\fi}
\def\nonvmodeerr@#1{\errmessage
    {\string#1\space is allowed only between paragraphs}}
\def\smallpagebreak{\relax\ifvmode\smallbreak
      \else\nonvmodeerr@\smallpagebreak\fi}
\def\medpagebreak{\relax\ifvmode\medbreak
       \else\nonvmodeerr@\medpagebreak\fi}
\def\bigpagebreak{\relax\ifvmode\bigbreak
      \else\nonvmodeerr@\bigpagebreak\fi}
\newdimen\captionwidth@
\captionwidth@=\hsize
\advance\captionwidth@ by -1.5in
\def\caption#1{}
\def\topspace#1{\gdef\thespace@{#1}\ifvmode\def\next
    {\futurelet\next\topspace@}\else\def\next{\nonvmodeerr@\topspace}\fi\next}
\def\topspace@{\ifx\next\Space@\def\next. {\futurelet\next\topspace@@}\else
     \def\next.{\futurelet\next\topspace@@}\fi\next.}
\def\topspace@@{\ifx\next\caption\let\next\topspace@@@\else
    \let\next\topspace@@@@\fi\next}
 \def\topspace@@@@{\topinsert\vbox to
       \thespace@{}\endinsert}
\def\topspace@@@\caption#1{\topinsert\vbox to
    \thespace@{}\nobreak
      \smallskip
    \setbox\z@=\hbox{\noindent\ignorespaces#1\unskip}%
   \ifdim\wd\z@>\captionwidth@
   \centerline{\vbox{\hsize=\captionwidth@\noindent\ignorespaces#1\unskip}}%
   \else\centerline{\box\z@}\fi\endinsert}
\def\midspace#1{\gdef\thespace@{#1}\ifvmode\def\next
    {\futurelet\next\midspace@}\else\def\next{\nonvmodeerr@\midspace}\fi\next}
\def\midspace@{\ifx\next\Space@\def\next. {\futurelet\next\midspace@@}\else
     \def\next.{\futurelet\next\midspace@@}\fi\next.}
\def\midspace@@{\ifx\next\caption\let\next\midspace@@@\else
    \let\next\midspace@@@@\fi\next}
 \def\midspace@@@@{\midinsert\vbox to
       \thespace@{}\endinsert}
\def\midspace@@@\caption#1{\midinsert\vbox to
    \thespace@{}\nobreak
      \smallskip
      \setbox\z@=\hbox{\noindent\ignorespaces#1\unskip}%
      \ifdim\wd\z@>\captionwidth@
    \centerline{\vbox{\hsize=\captionwidth@\noindent\ignorespaces#1\unskip}}%
    \else\centerline{\box\z@}\fi\endinsert}
\mathchardef\prime@="0230
\def\prime{{{}\prime@{}}}
\def\prim@s{\prime@\futurelet\next\pr@m@s}

\def\,{\relax\ifmmode\mskip\thinmuskip\else\thinspace\fi}
\def\!{\relax\ifmmode\mskip-\thinmuskip\else\negthinspace\fi}
\def\frac#1#2{{#1\over#2}}

\def\:{\nobreak\hskip.1111em{:}\hskip.3333em plus .0555em\relax}
\def\intic@{\mathchoice{\hskip5\p@}{\hskip4\p@}{\hskip4\p@}{\hskip4\p@}}
\def\negintic@
 {\mathchoice{\hskip-5\p@}{\hskip-4\p@}{\hskip-4\p@}{\hskip-4\p@}}
\def\intkern@{\mathchoice{\!\!\!}{\!\!}{\!\!}{\!\!}}
\def\intdots@{\mathchoice{\cdots}{{\cdotp}\mkern1.5mu
    {\cdotp}\mkern1.5mu{\cdotp}}{{\cdotp}\mkern1mu{\cdotp}\mkern1mu
      {\cdotp}}{{\cdotp}\mkern1mu{\cdotp}\mkern1mu{\cdotp}}}
\newcount\intno@
\def\iint{\intno@=\tw@\futurelet\next\ints@}
\def\iiint{\intno@=\thr@@\futurelet\next\ints@}
\def\iiiint{\intno@=4 \futurelet\next\ints@}
\def\idotsint{\intno@=\z@\futurelet\next\ints@}
\def\ints@{\findlimits@\ints@@}
\newif\iflimtoken@
\newif\iflimits@
\def\findlimits@{\limtoken@false\limits@false\ifx\next\limits
 \limtoken@true\limits@true\else\ifx\next\nolimits\limtoken@true\limits@false
    \fi\fi}
\def\multintlimits@{\intop\ifnum\intno@=\z@\intdots@
  \else\intkern@\fi
    \ifnum\intno@>\tw@\intop\intkern@\fi
     \ifnum\intno@>\thr@@\intop\intkern@\fi\intop}
\def\multint@{\int\ifnum\intno@=\z@\intdots@\else\intkern@\fi
   \ifnum\intno@>\tw@\int\intkern@\fi
    \ifnum\intno@>\thr@@\int\intkern@\fi\int}
\def\ints@@{\iflimtoken@\def\ints@@@{\iflimits@
   \negintic@\mathop{\intic@\multintlimits@}\limits\else
    \multint@\nolimits\fi\eat@}\else
     \def\ints@@@{\multint@\nolimits}\fi\ints@@@}
\def\Sb{_\bgroup\vspace@
        \baselineskip=\fontdimen10 \scriptfont\tw@
        \advance\baselineskip by \fontdimen12 \scriptfont\tw@
        \lineskip=\thr@@\fontdimen8 \scriptfont\thr@@
        \lineskiplimit=\thr@@\fontdimen8 \scriptfont\thr@@
        \Let@\vbox\bgroup\halign\bgroup \hfil$\scriptstyle
            {##}$\hfil\cr}
\def\endSb{\crcr\egroup\egroup\egroup}
\def\Sp{^\bgroup\vspace@
        \baselineskip=\fontdimen10 \scriptfont\tw@
        \advance\baselineskip by \fontdimen12 \scriptfont\tw@
        \lineskip=\thr@@\fontdimen8 \scriptfont\thr@@
        \lineskiplimit=\thr@@\fontdimen8 \scriptfont\thr@@
        \Let@\vbox\bgroup\halign\bgroup \hfil$\scriptstyle
            {##}$\hfil\cr}
\def\endSp{\crcr\egroup\egroup\egroup}
\def\Let@{\relax\iffalse{\fi\let\\=\cr\iffalse}\fi}
\def\vspace@{\def\vspace##1{\noalign{\vskip##1 }}}
\def\aligned{\,\vcenter\bgroup\vspace@\Let@\openup\jot\m@th\ialign
  \bgroup \strut\hfil$\displaystyle{##}$&$\displaystyle{{}##}$\hfil\crcr}
\def\endaligned{\crcr\egroup\egroup}
\def\matrix{\,\vcenter\bgroup\Let@\vspace@
    \normalbaselines
  \m@th\ialign\bgroup\hfil$##$\hfil&&\quad\hfil$##$\hfil\crcr
    \mathstrut\crcr\noalign{\kern-\baselineskip}}
\def\endmatrix{\crcr\mathstrut\crcr\noalign{\kern-\baselineskip}\egroup
                \egroup\,}
\newtoks\hashtoks@
\hashtoks@={#}
\def\format{\crcr\egroup\iffalse{\fi\ifnum`}=0 \fi\format@}
\def\format@#1\\{\def\preamble@{#1}%
  \def\c{\hfil$\the\hashtoks@$\hfil}%
  \def\r{\hfil$\the\hashtoks@$}%
  \def\l{$\the\hashtoks@$\hfil}%
  \setbox\z@=\hbox{\xdef\Preamble@{\preamble@}}\ifnum`{=0 \fi\iffalse}\fi
   \ialign\bgroup\span\Preamble@\crcr}

\def\cases{\left\{\,\vcenter\bgroup\vspace@
     \normalbaselines\openup\jot\m@th
       \Let@\ialign\bgroup$##$\hfil&\quad$##$\hfil\crcr
      \mathstrut\crcr\noalign{\kern-\baselineskip}}

\newif\iftagsleft@
\tagsleft@true
\def\TagsOnRight{\global\tagsleft@false}
\def\tag#1$${\iftagsleft@\leqno\else\eqno\fi
 \hbox{\def\pagebreak{\global\postdisplaypenalty-\@M}%
 \def\nopagebreak{\global\postdisplaypenalty\@M}\rm(#1\unskip)}%
  $$\postdisplaypenalty\z@\ignorespaces}
\interdisplaylinepenalty=\@M
\def\allowdisplaybreak@{\def\allowdisplaybreak{\noalign{\allowbreak}}}
\def\displaybreak@{\def\displaybreak{\noalign{\break}}}
\def\align#1\endalign{\def\tag{&}\vspace@\allowdisplaybreak@\displaybreak@
  \iftagsleft@\lalign@#1\endalign\else
   \ralign@#1\endalign\fi}
\def\ralign@#1\endalign{\displ@y\Let@\tabskip\centering\halign to\displaywidth
     {\hfil$\displaystyle{##}$\tabskip=\z@&$\displaystyle{{}##}$\hfil
       \tabskip=\centering&\llap{\hbox{(\rm##\unskip)}}\tabskip\z@\crcr
             #1\crcr}}
\def\lalign@
 #1\endalign{\displ@y\Let@\tabskip\centering\halign to \displaywidth
   {\hfil$\displaystyle{##}$\tabskip=\z@&$\displaystyle{{}##}$\hfil
   \tabskip=\centering&\kern-\displaywidth
        \rlap{\hbox{(\rm##\unskip)}}\tabskip=\displaywidth\crcr
               #1\crcr}}
\def\overrightarrow{\mathpalette\overrightarrow@}
\def\overrightarrow@#1#2{\vbox{\ialign{$##$\cr
    #1{-}\mkern-6mu\cleaders\hbox{$#1\mkern-2mu{-}\mkern-2mu$}\hfill
     \mkern-6mu{\to}\cr
     \noalign{\kern -1\p@\nointerlineskip}
     \hfil#1#2\hfil\cr}}}
\def\overleftarrow{\mathpalette\overleftarrow@}
\def\overleftarrow@#1#2{\vbox{\ialign{$##$\cr
     #1{\leftarrow}\mkern-6mu\cleaders\hbox{$#1\mkern-2mu{-}\mkern-2mu$}\hfill
      \mkern-6mu{-}\cr
     \noalign{\kern -1\p@\nointerlineskip}
     \hfil#1#2\hfil\cr}}}
\def\overleftrightarrow{\mathpalette\overleftrightarrow@}
\def\overleftrightarrow@#1#2{\vbox{\ialign{$##$\cr
     #1{\leftarrow}\mkern-6mu\cleaders\hbox{$#1\mkern-2mu{-}\mkern-2mu$}\hfill
       \mkern-6mu{\to}\cr
    \noalign{\kern -1\p@\nointerlineskip}
      \hfil#1#2\hfil\cr}}}
\def\underrightarrow{\mathpalette\underrightarrow@}
\def\underrightarrow@#1#2{\vtop{\ialign{$##$\cr
    \hfil#1#2\hfil\cr
     \noalign{\kern -1\p@\nointerlineskip}
    #1{-}\mkern-6mu\cleaders\hbox{$#1\mkern-2mu{-}\mkern-2mu$}\hfill
     \mkern-6mu{\to}\cr}}}
\def\underleftarrow{\mathpalette\underleftarrow@}
\def\underleftarrow@#1#2{\vtop{\ialign{$##$\cr
     \hfil#1#2\hfil\cr
     \noalign{\kern -1\p@\nointerlineskip}
     #1{\leftarrow}\mkern-6mu\cleaders\hbox{$#1\mkern-2mu{-}\mkern-2mu$}\hfill
      \mkern-6mu{-}\cr}}}
\def\underleftrightarrow{\mathpalette\underleftrightarrow@}
\def\underleftrightarrow@#1#2{\vtop{\ialign{$##$\cr
      \hfil#1#2\hfil\cr
    \noalign{\kern -1\p@\nointerlineskip}
     #1{\leftarrow}\mkern-6mu\cleaders\hbox{$#1\mkern-2mu{-}\mkern-2mu$}\hfill
       \mkern-6mu{\to}\cr}}}
\def\sqrt#1{\radical"270370 {#1}}
\def\dots{\relax\ifmmode\let\next=\ldots\else\let\next=\tdots@\fi\next}
\def\tdots@{\unskip\ \tdots@@}
\def\tdots@@{\futurelet\next\tdots@@@}
\def\tdots@@@{$\mathinner{\ldotp\ldotp\ldotp}\,
   \ifx\next,$\else
   \ifx\next.\,$\else
   \ifx\next;\,$\else
   \ifx\next:\,$\else
   \ifx\next?\,$\else
   \ifx\next!\,$\else
   $ \fi\fi\fi\fi\fi\fi}
\def\text{\relax\ifmmode\let\next=\text@\else\let\next=\text@@\fi\next}
\def\text@@#1{\hbox{#1}}
\def\text@#1{\mathchoice
 {\hbox{\everymath{\displaystyle}\def\textfonti{\the\textfont1 }%
    \def\textfontii{\the\textfont2 }\textdef@@ T#1}}
 {\hbox{\everymath{\textstyle}\def\textfonti{\the\textfont1 }%
    \def\textfontii{\the\textfont2 }\textdef@@ T#1}}
 {\hbox{\everymath{\scriptstyle}\def\textfonti{\the\scriptfont1 }%
   \def\textfontii{\the\scriptfont2 }\textdef@@ S\rm#1}}
 {\hbox{\everymath{\scriptscriptstyle}\def\textfonti{\the\scriptscriptfont1 }%
   \def\textfontii{\the\scriptscriptfont2 }\textdef@@ s\rm#1}}}
\def\textdef@@#1{\textdef@#1\rm \textdef@#1\bf
   \textdef@#1\sl \textdef@#1\it}

\def\textdef@#1#2{\def\next{\csname\expandafter\eat@\string#2fam\endcsname}%
\if S#1\edef#2{\the\scriptfont\next\relax}%
 \else\if s#1\edef#2{\the\scriptscriptfont\next\relax}%
 \else\edef#2{\the\textfont\next\relax}\fi\fi}
\scriptfont\itfam=\tenit \scriptscriptfont\itfam=\tenit
\scriptfont\slfam=\tensl \scriptscriptfont\slfam=\tensl
\mathcode`\0="0030
\mathcode`\1="0031
\mathcode`\2="0032
\mathcode`\3="0033
\mathcode`\4="0034
\mathcode`\5="0035
\mathcode`\6="0036
\mathcode`\7="0037
\mathcode`\8="0038
\mathcode`\9="0039
\def\Cal{\relax\ifmmode\let\next=\Cal@\else
     \def\next{\errmessage{Use \string\Cal\space only in math mode}}\fi\next}
\def\Cal@#1{{\fam2 #1}}
\def\bold{\relax\ifmmode\let\next=\bold@\else
   \def\next{\errmessage{Use \string\bold\space only in math
      mode}}\fi\next}\def\bold@#1{{\fam\bffam #1}}
\mathchardef\Gamma="0000
\mathchardef\Delta="0001
\mathchardef\Theta="0002
\mathchardef\Lambda="0003
\mathchardef\Xi="0004
\mathchardef\Pi="0005
\mathchardef\Sigma="0006
\mathchardef\Upsilon="0007
\mathchardef\Phi="0008
\mathchardef\Psi="0009
\mathchardef\Omega="000A
\mathchardef\varGamma="0100
\mathchardef\varDelta="0101
\mathchardef\varTheta="0102
\mathchardef\varLambda="0103
\mathchardef\varXi="0104
\mathchardef\varPi="0105
\mathchardef\varSigma="0106
\mathchardef\varUpsilon="0107
\mathchardef\varPhi="0108
\mathchardef\varPsi="0109
\mathchardef\varOmega="010A
\font\dummyft@=dummy
\fontdimen1 \dummyft@=\z@
\fontdimen2 \dummyft@=\z@
\fontdimen3 \dummyft@=\z@
\fontdimen4 \dummyft@=\z@
\fontdimen5 \dummyft@=\z@
\fontdimen6 \dummyft@=\z@
\fontdimen7 \dummyft@=\z@
\fontdimen8 \dummyft@=\z@
\fontdimen9 \dummyft@=\z@
\fontdimen10 \dummyft@=\z@
\fontdimen11 \dummyft@=\z@
\fontdimen12 \dummyft@=\z@
\fontdimen13 \dummyft@=\z@
\fontdimen14 \dummyft@=\z@
\fontdimen15 \dummyft@=\z@
\fontdimen16 \dummyft@=\z@
\fontdimen17 \dummyft@=\z@
\fontdimen18 \dummyft@=\z@
\fontdimen19 \dummyft@=\z@
\fontdimen20 \dummyft@=\z@
\fontdimen21 \dummyft@=\z@
\fontdimen22 \dummyft@=\z@
\def\fontlist@{\\{\tenrm}\\{\sevenrm}\\{\fiverm}\\{\teni}\\{\seveni}%
 \\{\fivei}\\{\tensy}\\{\sevensy}\\{\fivesy}\\{\tenex}\\{\tenbf}\\{\sevenbf}%
 \\{\fivebf}\\{\tensl}\\{\tenit}\\{\tensmc}}
\def\dodummy@{{\def\\##1{\global\let##1=\dummyft@}\fontlist@}}
\newif\ifsyntax@
\newcount\countxviii@
\def\newtoks@{\alloc@5\toks\toksdef\@cclvi}
\def\nopages@{\output={\setbox\z@=\box\@cclv \deadcycles=\z@}\newtoks@\output}
\def\syntax{\syntax@true\dodummy@\countxviii@=\count18
\loop \ifnum\countxviii@ > \z@ \textfont\countxviii@=\dummyft@
   \scriptfont\countxviii@=\dummyft@ \scriptscriptfont\countxviii@=\dummyft@
     \advance\countxviii@ by-\@ne\repeat
\dummyft@\tracinglostchars=\z@
  \nopages@\frenchspacing\hbadness=\@M}
\def\magstep#1{\ifcase#1 1000\or
 1200\or 1440\or 1728\or 2074\or 2488\or
 \errmessage{\string\magstep\space only works up to 5}\fi\relax}
{\lccode`\2=`\p \lccode`\3=`\t
 \lowercase{\gdef\tru@#123{#1truept}}}

\def\scaletype#1{\mag=#1\relax
 \hsize=\expandafter\tru@\the\hsize
 \vsize=\expandafter\tru@\the\vsize
 \dimen\footins=\expandafter\tru@\the\dimen\footins}

\def\scalefont#1#2\andcallit#3{\edef\font@{\the\font}#1\font#3=
  \fontname\font\space scaled #2\relax\font@}
\def\Mag@#1#2{\ifdim#1<1pt\multiply#1 #2\relax\divide#1 1000 \else
  \ifdim#1<10pt\divide#1 10 \multiply#1 #2\relax\divide#1 100\else
  \divide#1 100 \multiply#1 #2\relax\divide#1 10 \fi\fi}
\def\scalelinespacing#1{\Mag@\baselineskip{#1}\Mag@\lineskip{#1}%
  \Mag@\lineskiplimit{#1}}
\def\wlog#1{\immediate\write-1{#1}}
\catcode`\@=\active
\mag=\magstep1
\baselineskip=20pt
\hsize=16truecm
\vsize=23truecm
\pageno=1
\TagsOnRight
\def\=def{\; \mathop{=}_{\text{\rm def}} \;}
\def\del{\partial}

\def\Z{{\bold Z}}
\def\calL{{\cal L}}
\def\calU{{\cal U}}
\def\calV{{\cal V}}
\def\calLhat{{\hat{\calL}}}
\def\calUhat{{\hat{\calU}}}
\def\calVhat{{\hat{\calV}}}
\def\calT{{\cal T}}
\def\Lhat{{\hat{L}}}
\def\Vhat{{\hat{V}}}
\def\ellhat{{\hat{\ell}}}
\def\uhat{{\hat{u}}}
\def\vhat{{\hat{v}}}
\def\phat{{\hat{p}}}
\def\qhat{{\hat{q}}}
\def\phihat{{\hat{\phi}}}
\def\lambdahat{{\hat{\lambda}}}
%
\begingroup
\baselineskip=14pt
\line{{\it College of Liberal Arts and Sciences} \hfill KUCP-0046}
\line{{\it Kyoto University} \hfill hep-th/9203034}
\line{\hfill March, 1992}
\endgroup
\title
    Volume-preserving diffeomorphisms in \\
    integrable deformations of selfdual gravity \\
\endtitle
\author
    Kanehisa Takasaki\footnote"$^*$"{e-mail: TAKASAKI\@JPNRIFP.Bitnet}\\
    {\rm Institute of Mathematics, Yoshida College, Kyoto University}\\
    {\it Yoshida-Nihonmatsu-cho, Sakyo-ku, Kyoto 606, Japan}\\
\endauthor
\vglue2truecm
\heading
    Abstract
\endheading
\noindent
A group of volume-preserving diffeomorphisms in 3D turns out
to play a key role in an Einstein-Maxwell theory whose Weyl tensor
is selfdual and whose Maxwell tensor has algebraically general
anti-selfdual part. This model was first introduced by Flaherty
and recently studied by Park as an integrable deformation of selfdual
gravity.  A twisted volume form on the corresponding twistor space
is shown to be the origin of volume-preserving diffeomorphisms. An
immediate consequence is the existence of an infinite number of
symmetries as a generalization of $w_{1+\infty}$ symmetries in
selfdual gravity. A possible relation to Witten's 2D string theory
is pointed out.

\newpage

Selfdual gravity (= Ricci-flat K\"ahler geometry) has been
a laboratory of a number of recent attempts at higher
dimensional integrable field theories such as:
extensions of conformal or integrable field theories in 2D
[1][2][3][4], $N=2$ superstring theory [5], topological gravity [6][7]
etc. Classical integrability of selfdual gravity is due to
Penrose's twistor method (nonlinear graviton construction)[8].
A consequence of this method is the existence of an infinite
number of symmetries [9][10] whose underlying
structure is a loop group $: S^1 \to $ SDiff(2) of area-preserving
diffeomorphisms (on a plane). This should be considered a higher
dimensional extension of $w_{1+\infty}$ symmetries [11].

Selfdual gravity in the above narrow sense is by no means the only
possible integrable field theory of 4D gravity. Any conformally selfdual
space (Riemannian manifold with selfdual Weyl tensor) can be treated by
means of the nonlinear graviton construction, therefore may be thought
of as an ``integrable" field theory. Physical contents of conformally
selfdual spaces in general, however, are not very clear [12].

Flaherty [13] pointed out that scalar-flat K\"ahler (or ``quasi-K\"ahler"
in the terminology of Flaherty) geometry gives an interesting subfamily
of conformally selfdual spaces. According to Flaherty's observations,
these spaces are solutions of an Einstein-Maxwell theory whose Weyl
tensor is selfdual and whose Maxwell tensor has algebraically general
anti-selfdual part. Calling this system an ``integrable deformation"
of selfdual gravity, Park [14] presented a linear system that reproduces
the equation of motion (scalar-flatness) as the Frobenius integrability
condition.

We now show, in this letter, that Park's linear system has a
hidden SDiff(3) (= group of volume-preserving diffeomorphisms
in 3D) structure, and that this structure stems from a twisted
volume form on a corresponding 3D twistor space.  The existence
of SDiff(3) symmetries is an immediate consequence of these
facts.  The SDiff(2) loop group for selfdual gravity is a subgroup
of this SDiff(3) group, just as selfdual gravity is a special case
of Flaherty's Einstein-Maxwell theory with vanishing Maxwell fields.

Flaherty spaces, by definition, have a quasi-K\"ahler metric
$$
    ds^2 = \Omega_{,p^a \phat^b} dp^a d\phat^b, \quad
    p^a = (p,q), \
    \phat^a = (\phat,\qhat)                             \tag 1
$$
with a K\"aher potential $\Omega$ obeying the nonlinear equation
$$
    \Omega_{,p\phat} (\log D)_{,q\qhat}
  - \Omega_{,p\qhat}(\log D)_{,q\phat}
  - \Omega_{,q\phat} (\log D)_{,p\qhat}
  + \Omega_{,q\qhat}(\log D)_{,p\phat} = 0,             \tag 2
$$
where
$$
    D \=def \Omega_{,p\phat} \Omega_{,q\qhat}
            - \Omega_{,p\qhat} \Omega_{,q\phat} \not= 0.  \tag 3
$$
If $D=1$, Eq. (2) becomes an identity whereas Eq. (3) gives Plebanski's
heavenly equation for selfdual gravity [15]. It is convenient to
consider the following Poisson brackets.
$$
    \{ F, G\}_{pq} \=def F_{,p}G_{,q} - G_{,p}F_{,q},
    \quad
    \{ F, G\}_{\phat\qhat}
    \=def F_{,\phat}G_{,\qhat} - G_{,\phat}F_{,\qhat}.   \tag 4
$$
Eq. (2) then can be written
$$
    \{ \Omega_{,p}, (\log D)_{,q} \}_{\phat\qhat}
  - \{ \Omega_{,q}, (\log D)_{,p} \}_{\phat\qhat} = 0     \tag 5
$$
or, equivalently,
$$
    \{ \Omega_{,\phat}, (\log D)_{,\qhat} \}_{pq}
  - \{ \Omega_{,\qhat}, (\log D)_{,\phat} \}_{pq} = 0.    \tag 6
$$
Similarly, Eq. (3) have two equivalent expressions:
$$
    D = \{ \Omega_{,p}, \Omega_{,q} \}_{\phat\qhat}
      = \{ \Omega_{,\phat}, \Omega_{,\qhat} \}_{pq}.     \tag 7
$$
Eqs. (5) and (6) can be used to introduce two potentials $\phi$
and $\phihat$:
$$
  \align
  & \phihat_{,p} = \{ \Omega_{,p}, \log D \}_{\phat\qhat},        \quad
    \phihat_{,q} = \{ \Omega_{,q}, \log D \}_{\phat\qhat}, \tag 8 \cr
  & \phi_{,\phat} = \{ \Omega_{,\phat}, \log D \}_{pq},           \quad
    \phi_{,\qhat} = \{ \Omega_{,\qhat}, \log D \}_{pq}.    \tag 9 \cr
  \endalign
$$
The four functions $(\Omega,D,\phi,\phihat)$ are the fundamental
ingredients of this model.

Park's linear system, in our notation, can be written
$$
    L_p \psi = 0,  \quad  L_q \psi = 0,                   \tag 10
$$
where
$$
  \align
  &  L_p
     \=def \del_p + \lambdahat\{ \Omega_{,p}, \cdot\}_{\phat\qhat}
           -\phihat_{,p} \lambdahat^2 \del_\lambdahat,                \cr
  &  L_q
     \=def \del_q + \lambdahat\{ \Omega_{,q}, \cdot\}_{\phat\qhat}
           -\phihat_{,q} \lambdahat^2 \del_\lambdahat,        \tag 11 \cr
  \endalign
$$
and $\lambdahat$ is a new variable (``spectral parameter" in selfdual
gravity). The equations of motion for $\Omega$ and $D$ imply the
integrability condition $ [L_p, L_q] = 0$ of this linear system
(and vice versa).  In fact, since the roles of $(\phi,p,q)$ and
$(\phihat,\phat,\qhat)$ are interchangeable, one can introduce
another (dual) linear system:
$$
    \Lhat_\phat \psi = 0,   \quad \Lhat_\qhat \psi = 0,      \tag 12
$$
where
$$
  \align
  &  \Lhat_\phat
      \=def \del_\phat - \lambda^{-1}\{ \Omega_{,\phat}, \cdot \}_{pq}
            - \phi_{,\phat}\del_\lambda,                             \cr
  &  \Lhat_\qhat
      \=def \del_\qhat - \lambda^{-1}\{ \Omega_{,\qhat}, \cdot \}_{pq}
            - \phi_{,\qhat}\del_\lambda,                  \tag 13    \cr
  \endalign
$$
and
$$
    \lambda \=def D \lambdahat.                           \tag 14
$$
This linear system, too, is integrable: $[\Lhat_\phat, \Lhat_\qhat] = 0$.

The Frobenius integrability in this case means that each of the above
linear systems has exactly three functionally independent solutions,
say $(\psi_0,\psi_1,\psi_2)$ with
$ d\psi_0 \wedge d\psi_1 \wedge d\psi_2 \not= 0$.
Let us call such a triple a set of fundamental solutions.
It should be noted that any 3D diffeomorphism
$f = (f_0,f_1,f_2)$ can act on fundamental solutions to
give a new set of fundamental solutions:
$$
    (\psi_0,\psi_1,\psi_2) \to
    \bigl( f_0(\psi_0,\psi_1,\psi_2),
           f_1(\psi_0,\psi_1,\psi_2),
           f_2(\psi_0,\psi_1,\psi_2) \bigr).             \tag 15
$$
A clue of the twistor method is to consider two sets of special
fundamental solutions, $(\calL,\calU,\calV)$ and
$(\calLhat,\calUhat,\calVhat)$, with different analyticity.
Namely, they are required to be holomorphic functions of $\lambda$
(or of $\lambdahat$) in a neighborhood of a circle, say
$|\lambda|=r$, and to have analytic continuation outside
(for $\calL,\calU,\calV$) or inside (for $\calLhat,\calUhat,\calVhat$)
that circle with Laurent expansion of the following form:
$$
  \align
  \calL    = \lambda + \phi + \sum_{n=-\infty}^{-1} \ell_n \lambda^n, \quad&
  \calLhat = \lambdahat + \phihat \lambdahat^2
                        + \sum_{n=3}^\infty \ellhat_n \lambdahat^n,   \cr
  \calU    = p\calL + \sum_{n=-\infty}^{0} u_n \calL^n,               \quad&
  \calUhat = \phat + \sum_{n=1}^\infty \uhat_n \calLhat^n,            \cr
  \calV    = q\calL + \sum_{n=-\infty}^{0} v_n \calL^n,               \quad&
  \calVhat = \qhat + \sum_{n=1}^\infty \vhat_n \calLhat^n.   \tag 16  \cr
  \endalign
$$
According to the aforementioned general property of fundamental
solutions, these two sets of solutions should be linked by an
invertible functional relation:
$$
    \calLhat = f_0(\calL,\calU,\calV), \quad
    \calUhat = f_1(\calL,\calU,\calV), \quad
    \calVhat = f_2(\calL,\calU,\calV).                   \tag 17
$$
This is a kind of Riemann-Hilbert problem related to a group of
3D diffeomorphisms; the functional data $f=(f_0,f_1,f_3)$ is indeed
such a diffeomorphism.  Penrose's twistor method is a geometric
interpretation of this Riemann-Hilbert problem in the language
of a curved twistor space $\calT$.

Actually, our linear systems have a special structure that allows us
to put a set of constraints on $(\calL,\calU,\calV)$ and
$(\calLhat,\calUhat,\calVhat)$. To see this, note that the
$(\del_\lambda,\del_p,\del_q)$ part of $\Lhat_\phat$ and $\Lhat_\qhat$
in (12) are divergence-free vector fields on the $(\lambda,p,q)$ space,
i.e., of the form
$$
    V = v_0 \del_\lambda + v_1 \del_p  + v_2 \del_q,    \quad
    v_{0,\lambda} + v_{1,p} + v_{2,q} = 0.              \tag 18
$$
Similarly, the $(\del_\lambdahat,\del_\phat,\del_\qhat)$ part of
$L_p$ and $L_q$ in (10) are vector fields of the form
$$
  \align
  &  \Vhat = -\vhat_0 \lambdahat^2 \del_\lambdahat
             + \vhat_1 \del_\phat + \vhat_2 \del_\qhat
           = \vhat_0 \del_{\lambdahat^{-1}}
             + \vhat_1 \del_\phat + \vhat_2 \del_\qhat,
                                                                      \cr
  & - \lambdahat^2 \vhat_{0,\lambdahat} + \vhat_{1,\phat}
    + \vhat_{2,\qhat} = 0,                              \tag 19       \cr
  \endalign
$$
thereby divergence-free in $(\lambdahat^{-1},\phat,\qhat)$.
Both the linear systems, thus, may be thought of as defining
a commuting pair of 3D volume-preserving diffeomorphisms
[in the $(\lambda,p,q)$ space for Eq. (12) and
in the $(\lambdahat^{-1},\phat,\qhat)$ space for Eq. (10),
respectively]. Because of this, the fundamental solutions
$(\calL,\calU,\calV)$ and $(\calLhat,\calUhat,\calVhat)$ may
be selected to satisfy the following SDiff(3) constraints:
$$
  \align
    \frac{\del(\calL^{-1},\calU,\calV)}
          {\del(\lambdahat^{-1},\phat,\qhat)}
    = \frac{\del(\calLhat^{-1},\calUhat,\calVhat)}
          {\del(\lambdahat^{-1},\phat,\qhat)}
  & = 1,                                                \tag 20        \cr
    \frac{\del(\calL,\calL^{-1}\calU,\calL^{-1}\calV)}
          {\del(\lambda,p,q)}
    = \frac{\del(\calLhat,\calLhat^{-1}\calUhat,\calLhat^{-1}\calVhat)}
          {\del(\lambda,p,q)}
  & = 1.                                                \tag 21        \cr
  \endalign
$$

These constraints, along with the linear systems themselves,
can be cast into a more compact form.  One can indeed prove that
they altogether are equivalent to the exterior differential equation
$$
    d\calL^{-1} \wedge d\calU \wedge d\calV
    = \theta
    = d\calLhat^{-1} \wedge d\calUhat \wedge d\calVhat,   \tag 22
$$
where $\theta$ is the 3-form given by
$$
  \align
  \theta
  & = -d\lambda \wedge (dp \wedge dq + \lambda^{-1}\omega
         + \lambda^{-2} D d\phat \wedge d\qhat)
      -d\phi \wedge dp \wedge dq
      -\lambda^{-1}dD \wedge d\phat \wedge d\qhat                 \cr
  & = -\lambdahat^{-2} d\lambdahat \wedge ( d\phat \wedge d\qhat
         + \lambdahat \omega + \lambdahat^2 D dp \wedge dq )
      -d\phihat \wedge d\phat \wedge d\qhat
      -\lambdahat dD \wedge dp \wedge dq,                 \tag 23 \cr
  \endalign
$$
and $\omega$ is the K\"ahler 2-form,
$$
    \omega = \Omega_{,p^a \phat^b} dp^a \wedge d\phat^b.  \tag 24
$$
Geometrically, Eq. (22) means the existence of a twisted volume form
on the corresponding twistor space $\calT$. If, in particular,
$\calL=\calLhat = \lambda = \lambdahat$ (hence $D=1$), Eq. (22)
reduces to the 2-form equation of selfdual gravity [9][10].

The data $f = (f_0,f_1,f_2)$ of the Riemann-Hilbert problem is
also required to satisfy an SDiff(3) constraint of the form
$$
    \frac{\del(f_0^{-1},f_1,f_2)}{\del(\calL^{-1},\calU,\calV)} = 1.
                                                          \tag 25
$$
Conversely, solving the Riemann-Hilbert problem under this condition, we
can obtain in principle all Flaherty spaces. If $f$ leaves the first
component invariant as $f_0(\calL,\calU,\calV) = \calL$, the above
SDiff(3) condition becomes an SDiff(2) condition defining a loop group
element with values in SDiff(2), and we have in turn a solution of
selfdual gravity.

Now the existence of SDiff(3) symmetries is obvious; they are generated
by the action of the SDiff(3) group on the Riemann-Hilbert data $f$
from left or right.  Via the Riemann-Hilbert factorization, these
actions on $f$ give rise to actions on $(\calL,\calU,\calV)$
and $(\calLhat,\calUhat,\calVhat)$. Although finite symmetries are
very complicate in general, infinitesimal symmetries should take
a very simple and universal form as we have found for selfdual
gravity [1][10]. This issue will be reported elsewhere.

We have thus observed that an SDiff(3) structure, exhibited most clearly
by Eq. (22), is a key to characterize Flaherty spaces among general
conformally selfdual spaces. The theory turns out to be integrable
in the sense of the nonlinear graviton construction, and includes
selfdual gravity as a special case. The existence of SDiff(3) symmetries
can be deduced from these results.

It is amusing to compare our results with a recent observation
of Witten [16] on the relevance of an SDiff(3) group in 2D string
theory.  A fundamental volume form arising therein is written
$$
    \frac{da_1 \wedge da_2 \wedge da_3}{a_3}
    = da_1 \wedge da_2 \wedge d \log a_3.              \tag 26
$$
This strongly suggests that Witten's observation will be related
to a 3-form equation of the form
$$
      d\log\calL \wedge d\calU \wedge d\calV
    = d\log\calLhat \wedge d\calUhat \wedge d\calVhat  \tag 27
$$
in our notation. This 3-form equation, too, gives an integrable
deformation of self-dual gravity with SDiff(3) symmetries.
Remarkably, a similar equation (of 2-forms) including
$\log\calL$ and $\log\calLhat$ has been discovered in a study of a
continuous analogue of the Toda chain field theory [17]; this model
is shown to be related to the SDiff(2) group on a cylinder.

\heading
    Acknowledgements
\endheading
\noindent
The author would like to thank Q-Han Park and Jean Avan
for discussion and useful comments.

\newpage

\heading
    References
\endheading

\item{[1]}
Park, Q-Han,
Phys. Lett. B236 (1990), 429-432;
Phys. Lett. B238 (1990), 287-290;
Phys. Lett. B257 (1991), 105-110.

\item{[2]}
Kashaev, R.M., Saveliev, M.V., Savelieva, S.A., and Vershik, A.M.,
Institute for High Energy Physics, Moscow, preprint 90-I (1990).

\item{[3]}
Yamagishi, K., and Chapline, F.,
Class. Quantum Grav. 8 (1991), 427-446.

\item{[4]}
Hull, C.M.,
Phys. Lett. B269 (1991), 257-263.

\item{[5]}
Ooguri, H., and Vafa, C.,
Mod. Phys. Lett. A5 (1990), 1389-1398;
Nucl. Phys. B361 (1991), 469-518.

\item{[6]}
Oda, I., and Sugamoto, A.,
Phys. Lett. B266 (1991), 280-284.

\item{[7]}
Kunitomo, H.,
Mod. Phys. Lett. A6 (1991), 2389-2396.

\item{[8]}
Penrose, R.,
Gen. Rel. Grav. 7 (1976), 31-52.

\item{[9]}
Boyer, C.P., and Plebanski, J.F.,
J. Math. Phys. 26 (1985), 229-234.

\item{[10]}
Takasaki, K.,
J. Math. Phys. 31 (1990), 1877-1888.

\item{[11]}
Bakas, I.,
Phys. Lett. B228 (1989), 57-63;
Commun. Math. Phys. 134 (1990), 487-508.

\item{[13]}
Flaherty, E.J.,
Gen. Rel. Grav. 9 (1978), 961-978.

\item{[12]}
Boyer, C.P., and Plebanski, J.E.,
Phys. Lett. A106 (1984), 125-129.

\item{[14]}
Park, Q-Han,
Phys. Lett. B269 (1991), 271-274.

\item{[15]}
Plebanski, J.F.,
J. Math. Phys. 16 (1975), 2395-2402.

\item{[16]}
Witten, E.,
Institute for Advanced Study preprint IASSNS-HEP-91/51 (August, 1991).

\item{[17]}
Takasaki, K., and Takebe, T.,
Kyoto University preprint RIMS-790 (August, 1991),
Lett. Math. Phys. 23 (to appear).

\bye